\documentclass[conference]{IEEEtran}
\usepackage[utf8]{inputenc}%(only for the pdftex engine)

\usepackage{cleveref}
\usepackage{microtype}
\usepackage[dvipsnames]{xcolor}

\usepackage{tabularx}
\usepackage{csquotes}
\usepackage[inline]{enumitem}
\usepackage{flushend}

\usepackage{graphicx}
\DeclareGraphicsExtensions{.pdf,.png,.jpg}

\usepackage{url}

\ifCLASSINFOpdf
\else
\fi
\begin{document}
	\title{Exploratory Study of the Privacy Extension for System Theoretic Process Analysis (STPA-Priv) to elicit Privacy Risks in eHealth}
	
	\author{\IEEEauthorblockN{Kai Mindermann, Frederik Riedel, Asim Abdulkhaleq, Christoph Stach, Stefan Wagner}
		\IEEEauthorblockA{University of Stuttgart\\
			Universitätsstraße 38, 70569 Stuttgart}
	}
	
	
	%Telephone: (800) 555--1212, Fax: (888) 555--1212}


	\maketitle
	
	\begin{abstract}
		\emph{Context:} System Theoretic Process Analysis for Privacy (STPA-Priv) is a novel privacy risk elicitation method using a top down approach. It has not gotten very much attention but may offer a convenient structured approach and generation of additional artifacts compared to other methods.
		\emph{Aim:} The aim of this exploratory study is to find out what benefits the privacy risk elicitation method STPA-Priv has and to explain how the method can be used.
		\emph{Method:} Therefore we apply STPA-Priv to a real world health scenario that involves a smart glucose measurement device used by children. Different kinds of data from the smart device including location data should be shared with the parents, physicians, and urban planners. This makes it a sociotechnical system that offers adequate and complex privacy risks to be found.
		\emph{Results:} We find out that STPA-Priv is a structured method for privacy analysis and finds complex privacy risks. The method is supported by a tool called XSTAMPP which makes the analysis and its results more profound. Additionally, we learn that an iterative application of the steps might be necessary to find more privacy risks when more information about the system is available later.
		\emph{Conclusions:} STPA-Priv helps to identify complex privacy risks that are derived from sociotechnical interactions in a system. It also outputs privacy constraints that are to be enforced by the system to ensure privacy. 
	\end{abstract}
	

	% \fi
	\IEEEpeerreviewmaketitle

	\let\thefootnote\relax\footnotetext{This is the author's version of the work published at the Requirements Engineering Conference Workshops (REW), 2017 IEEE 25th International (\url{https://doi.org/10.1109/REW.2017.30}). \\ \copyright~2017 IEEE. Personal use of this material is permitted. Permission from IEEE must be obtained for all other users, including reprinting/ republishing this material for advertising or promotional purposes, creating new collective works for resale or redistribution to servers or lists, or reuse of any copyrighted components of this work in other works.}
	
	\section{Introduction}
	The increasing importance of privacy is relevant for organizations and individuals in a connected world. Especially for upcoming electronic health systems that are enabled by Internet of Things devices.
	Such complex socio-technical software systems offer personalized services, personal assistants, and cloud services. Collecting and processing personal information is essential for those services. According to a Gartner forecast \cite{Middleton13} the Internet of Things, as in everything from your toothbrush over your freezer to your television is connected to the Internet, will encompass 26 billion devices by 2020. Each of these devices will have different sensors ranging from a camera, microphone, and GPS sensors to more unfamiliar ones like motion, temperature and light sensors (and even more specific ones). As multiple devices have these sensors and they send the collected data usually to at least one service provider it is unknown on what basis this data is analyzed and related to other information. The research project PATRON (Privacy in Stream Processing) funded by the Baden-Württemberg Stiftung works on concealing privacy relevant patterns in data streams. It is important to find relevant privacy vulnerabilities to derive patterns. Previous analysis techniques appear to be non-systematic when it comes to the combination of data and interaction with the environment. For example with multiple data receivers that are not part of the original system. Safety and security can be considered a system property and system-theoretic approaches are being used as alternatives to established methods in their field. Privacy can also be seen as a system property as privacy relevant data can be leaked at all abstraction levels. Therefore we expect that system-theoretic methods can be applied with good results to privacy as well.
	The currently proposed STPA-Priv method by Shapiro \cite{Shapiro16} has not been getting a lot of attention and we want to explore if it is feasible to use the method for privacy analysis.
	
	\section{Related Work}
	Data-flow analysis techniques have been developed to track data flow and elicit privacy risks. The approach described by Lu and Li \cite{LuLi} includes different existing data-flow analysis-techniques such as "conditional flow identification" and "joint flow tracking". They implemented a system that analyzes Android application-files for malicious data-flow. This includes revealing contacts, call logs, browser history, Short Message Service (SMS) history, Global Positioning System (GPS), or unique user IDs. A similar system for iOS applications has been developed by Egele and Krueger \cite{Egele11}. Their system is able to detect data-flow in compiled Objective-C binaries, similar to Lu and Li's approach. Another interesting approach has been developed by Enck and Gilbert \cite{Enck14}. Their system can analyze data-flow in Android applications in real-time, in contrast to the static approach of Lu, Li, Egele, and Krueger. Their system TaintDroid can be run on productive devices in the background to spot malicious app requests.
	
	Analyzing data flow, such as suggested by Lu and Li \cite{LuLi}, Egele and Krueger \cite{Egele11}, and Enck and Gilbert \cite{Enck14}, focuses on data sharers, data observers and data exchanged between them. However, these three approaches are optimized for mobile applications and only consider access to initial information sources, such as contact information but do not elicit privacy risks that can occur with data that has been exchanged with other systems or participants.
	They do not consider what happens with these information outside of their scope. In many cases, it is necessary to exchange information for a service to be able to work as expected. The revealing of privacy information is not always a privacy risk. Later on when data is exchanged with other partners or combined with other data sets privacy risks can occur as well which would not be covered by these approaches.
	
	Another example for data-flow analysis is the LINDDUN methodology, described by Wuyts, Scandariato and Joosen \cite{LINDDUN14}. Their approach uses a data flow diagram as a starting point to find privacy threats. A privacy threat catalog is then used to categorize each entity of the diagram into seven possible threat categories: \textit{linkability}, \textit{identifiability}, \textit{non-repudiation}, \textit{detectability}, \textit{information disclosure}, \textit{unwareness}, and \textit{non-compliance} (hence the name LINDDUN). The model goes even further and describes a process to resolve privacy threats. 
	However, it is difficult to analyze complicated socio-technical systems using this approach, because it is focused on a bottom-up data flow analysis. Privacy risks often result from human interaction or interactions with different systems which makes it difficult for bottom-up analysis techniques to unveil these. Indeed, this is a drawback of this approach that has been proven by Wuyts, Scandariato, and Joosen in a set of extensive empirical studies \cite{Wuyts14}. They also state that it \enquote{[\ldots] mainly  focuses  on the privacy of the data subject (i.e.  the person the data are about). Rather than focusing on internal processes and flows[\ldots]} \cite{wuyts2015linddun}.
	This is where STPA-Priv could be useful with its top-down approach.

	\section{STPA-Priv}
	To understand what STPA-Priv is we first describe its origin and what ideas constituted to the development of STPA. Then we introduce the STPA-Priv extension and available tool support.

	\begin{table*}[htb]
		\renewcommand{\arraystretch}{1.4}
		\begin{tabularx}{\textwidth}{r|l|l|X}
			~                          & \textbf{STPA-Priv}                                & STPA-Sec                                     & STPA                                                \\ \hline
			\emph{Step 0} Fundamentals  & \multicolumn{3}{l}{Define System Goals and Description}                                            \\ \cline{2-4}
			~                          & Define \textbf{Adverse Consequences}              & Define Losses                                       &  Define Accidents                                    \\ \cline{2-4}
			~                          & \multicolumn{2}{l|}{Define Vulnerabilities}                                    & Define Hazards \\ \cline{2-4}
			~                          & Link Adverse Consequences to Vulnerabilities      & Link Losses to Vulnerabilities        & Link Accidents to Hazards      \\ \cline{2-4}
			~                          & Specify \textbf{Privacy Constraints}                      & Specify Security Constraints                         & Specify Safety Constraints                                     \\ \cline{2-4}
			~                          & \multicolumn{3}{l}{Specify Design Requirements}                                      \\ \cline{2-4}
			~                          & \multicolumn{3}{l}{Create Control Structure Model}                       \\ \hline
			~ & \multicolumn{3}{c}{~} \\
			\emph{Step 1} Control Actions           & \multicolumn{3}{l}{Derive Control Actions}  \\ \cline{2-4}
			~                          & Define \textbf{Privacy-compromising Control Actions} & Define Unsecure Control Actions & Define Unsafe Control Actions \\ \cline{2-4}
			~                          & Corresponding Privacy Constraints           	      & Corresponding Security Constraints           & Corresponding Safety Constraints                        \\ \hline
			~ & \multicolumn{3}{c}{~} \\
			\emph{Step 2} Causal Analysis          & \multicolumn{3}{l}{Derive Causal Factors} \\ 
		\end{tabularx}
		\caption{Overview of the steps of the STPA-Priv method with comparison to the original STPA and STPA-Sec.}
		\label{tab:STPA-Comparison}
	\end{table*}
	
	\subsection{STAMP and STPA}
	Leveson developed a new accident model based on system and control theory called STAMP (Systems-Theoretic Accident Modeling and Processes) \cite{Leveson11}. In STAMP, accidents are considered results from inadequate enforcement of safety constraints in system design, development and operations. STAMP treats safety as control problem rather than component failures. In STAMP, the system is seen as a set of control components which interact with each other. This helps to create models of systems which cover human, technology, software, and environmental factors, such as governmental policy \cite{Leveson11}. Therefore, STAMP considers accidents not only arising from individual component failures but also from the interaction among system components. In other words, accidents occur when component failures, external disturbances and/or dysfunctional interactions among system components are not adequately handled by the safety control system \cite{Leveson11}. Based on STAMP, a new method for hazard analysis called STPA (System-Theoretic Process Analysis) was developed to identify hazards existing in the system and providing so-called safety constraints to mitigate those hazards.
	
	STPA has been extended to support the security analysis based on systems theory. Young and Leveson \cite{Young} developed an approach called STPA-Sec (STPA for security) which extends the STPA safety analysis with security aspects.
	
	\subsection{STPA-Priv Extension}
	Shapiro extended STPA-Sec to be used for the elicitation of privacy risks \cite{Shapiro16}. His proposed extension is called \textbf{STPA-Priv}. It combines the existing advantages of STPA with an extension for privacy analysis. This includes the top-down principle of STPA to be able to handle complex socio-technical systems. The steps of STPA are in principle the same in STPA-Priv, only their terminology was changed. 
	Losses or accidents in traditional STPA are always related to a loss of human life, injuries or destruction of expensive hardware. 
	However, privacy violations primarily do not lead to accidents which threaten human life\footnote{Privacy violations can still lead to political or other kinds of persecution and should therefore not be neglected.}, but lead to embarrassing, awkward and adverse situations, or emotional damage in general, for individuals. This is why losses and accidents are renamed to \emph{adverse consequences} in respect to privacy. An important property of STPA-Priv is that it can cope with open-loop controls, that is controls that are not able to provide feedback to their controlling entity (e. g. privacy policies are there but can not alway be known if the user really read them \cite{Nagrath06}).
	An overview of the needed steps and comparison between STPA-Priv, STPA-Sec, and STPA is depicted in \Cref{tab:STPA-Comparison}. 
	
	Wuyts and Joosen define four different knowledge classes for privacy research \cite{Hazeyama2016}. They are \emph{methodology/process}, \emph{Principle}, \emph{Guideline} and \emph{Pattern}. Currently we would  classify STPA-Priv as a \emph{methodology/process}. But as we will learn in this exploratory study, it has to be augmented by at least a threat catalog for a useful analysis so it might also belong to the \emph{pattern} category.
	
%	\begin{figure*}[htb]
%		\includegraphics[width=\textwidth]{img/stpa-priv-process.png}
%		\caption{Steps in STPA-Priv and examples for results of each step. It starts with identifying \emph{adverse consequences}, continues with \emph{vulnerable system states}, deriving \emph{privacy constraints}, creating the \emph{control structure diagram}, identifying \emph{privacy compromising control actions} and finally generating \emph{causal scenarios}.}
%		\label{fig:stpa-priv-tasks}
%	\end{figure*}
	
	\subsection{Tool Support with XSTAMPP}
	The application of a method is easier with a tool that gives hints or makes the input/output of artifacts easier. XSTAMPP as open-source platform supports engineers to perform safety, security and privacy analysis by providing a mostly tabular user interface for each STPA step and reference capabilities between the artifacts. It also has graphical control structure modeling. XSTAMPP supports software engineers in performing safety-based testing and verification activities based on the STPA safety analysis results. We will use XSTAMPP to support our application of STPA-Priv on a reasonable complex scenario which we will describe next. 
	
	\section{Scenario Description (eHealth)}
	Since chronic diseases such as diabetes mellitus are on the rise, the healthcare system has to face high treatment costs and overburdened physicians. As a consequence, novel treatment methods are badly needed. eHealth, i.\,e., the usage of common computing systems such as PCs or smartphones for health care, is such a method. With eHealth the patients are able to perform periodic screenings at home instructed only by their computing system. eHealth applications can even be tailored to almost any given medical condition~\cite{Siewiorek2012}. However, eHealth is not just able to reduce treatment costs. With the help of \emph{serious games} it is possible to integrate required therapeutic procedures into the daily routines of child patients. That way, the young patients are able to cope with their illness much better~\cite{Knoell2009}.
	
	Knöll develops in cooperation with the Olgahospital Stuttgart, a children's hospital in Germany, an idea for a serious game for children suffering from diabetes~\cite{Knoell2010}. In this smartphone game the player, i.\,e., the patient, has to enter his or her blood sugar values regularly. Each of these entries is enriched with the current location of the player and a timestamp\footnote{Both, the location as well as the current time can be captured by the sensors built in the smartphone.}. The game somehow has to motivate the player not only to check his or her blood sugar level regularly but also to do this at varying locations. From the player's point of view, this concept is beneficial as s/he has to monitor the blood sugar level anyhow. Accordingly the game is both a motivation as well as a reminder. Additionally physicians benefit from such a game. Normally patients have to keep a handwritten diabetes diary and the physicians have to review the diary. Yet, these diaries contain wrong or incomplete data and they are difficult to decipher. The game-based approach is able to create an electronic diabetes diary without the drawbacks of the paper-based version.
	
	However, there are even more stakeholders for such a game. Due to the augmented health data, also urban planners can profit from health games. With the help of physicians, they are able to identify unhealthy places in town, i.\,e., places which have a bad influence on the patient's health. Based on this knowledge, correlations between architectural characteristics (e.\,g., crowded streets) and health condition changes can be deduced~\cite{Knoell2014a}.
	
	It is obvious that not every involved party requires all of the captured data. Especially since this kind of data is highly sensitive private data. Therefore it is strongly recommended to restrict the party's data access individually. E.\,g., the urban planners only need to know which locations have an influence on the health condition. However, they need no access to any actual health data or even data from which they are able to draw inferences about the patient. The physicians on the other hand need access to the whereabouts of a patient only in case of an emergency. Such a scenario requires fine-granular privacy mechanisms which assure the best quality of service and conceal as much private data as possible. 
	The studies mentioned above have not taken privacy into consideration; the participants had to agree to the unrestricted usage of their data by the involved parties. We assume that this scenario covers several different privacy risks that can be found in similar eHealth scenarios as well.

	\section{Application of STPA-Priv}
 	In the following we want to systematically analyze this scenario using STPA-Priv for any privacy risks the involved parties have to be aware of and how these risks can be mitigated or even prevented. The analysis follows the steps listed in \Cref{tab:STPA-Comparison}.  
	
	\subsection*{Step 0: Fundamentals}
	
	\subsubsection{Define System Goals and Description}
	Important goals of the system have to be kept in mind in all further steps, as the system must always fulfill its goal. 
	We can get important information from the system description, such as the involved parties that share or process data. In this scenario we come up with the following initial involved parties: 
	\emph{Child} (as in the person that has diabetes), \emph{parents}, \emph{physician}, \emph{insurance company}, \emph{smart device manufacturer}, and \emph{other players} of the game (other children with diabetes).
	
	\subsubsection{Define Adverse Consequences}
	Finding and defining \textbf{adverse consequences} in our system is an important step of STPA-Priv and requires experts that know the scenario and its entities. However, it is not necessary to know the implementation of each component, since STPA is a top-down approach and works at the system and component level. 
	
	This step can and should be augmented by a systematic catalog of privacy threats, such as \textit{LINDDUN privacy threat tree catalog} \cite{LINDDUN14} or \emph{Calo’s subjective/objective privacy harms} \cite{Calo11} (as Shapiro used in his initial proposal of STPA-Priv \cite{Shapiro16}). LINDDUN has been analyzed in empirical studies that tested how different threat models affect the traceability of different privacy threats. These studies showed that this threat model is easy to learn but still provides reliable results in comparison to experts \cite{Wuyts14}. Its threat trees have been considered useful in practice. It provides privacy analysis methods as well, however, we only utilized their threat tree in our case. It offers privacy threats from different categories: \textit{linkability}, \textit{identifiability}, \textit{non-repudiation}, \textit{detectability}, \textit{information disclosure}, \textit{unwareness}, and \textit{non-compliance}. Threats from these categories are then used to find adverse consequences. 
	
	Each adverse consequence can be triggered by one or more system states together with environmental conditions of the system. These are called \textbf{vulnerabilities or vulnerable system states}. Vulnerabilities are system states that are under the system’s control, whereas adverse consequences themselves are not controllable. This is why vulnerable system states have to be prevented. Elaborating adverse consequences is the counterpart to accidents in original STPA. Here we can use the knowledge from the previous step about the involved parties and their relationships.
	
	As an example, the relationship between the child and the smart device manufacturer is of commercial interest. Exchanged data includes analytics data and crash report information. Applying different privacy threats from LINDDUN threat tree create the following adverse consequence: \emph{The user is not aware of active analytics program and is therefore suspect to surveillance}. This is a result of the general privacy threat \textit{unawareness}. Another example is \emph{other players can estimate health state of player} which is caused by \emph{information disclosure} \cite{Shapiro16}. More adverse consequences are listed in \cref{tab:AdverseConsequences}.
	
	\subsubsection{Define Vulnerabilities}
	Now that we have a list of adverse consequences we need to define corresponding \textbf{vulnerable system states} that can lead to these adverse consequences \cite{Shapiro16} \cite{Leveson11}. 
	Depending on the adverse consequence we can define an abstract system state description for each adverse consequence that would be exploitable, respectively that can lead to the adverse consequence.
	At this point we have no list or model of possible system states. Therefore, we assume we have to describe them in a textual form with our domain knowledge and the knowledge of the system and they do not have to be actual states in the implemented components of the system.
	
	The adverse consequence \emph{the user is not aware of active analytics program and is therefore suspect to surveillance} can be caused by the system states \textit{privacy policy has not been presented to user} and \emph{user ignored privacy policy and did not read it}. A list of identified vulnerable system states of our scenario is listed in \cref{tab:AdverseConsequences} where we also state the adverse consequence and the LINDDUN category.
	
	\begin{table*}
		\renewcommand{\arraystretch}{1.2}
		\begin{tabularx}{\textwidth}{XlX}
			Adverse Privacy Consequences                                                            & LINDDUN Category                    & Vulnerable System States                                                                                        \\ \hline
			User is not aware of active analytics program and is therefore suspect to surveillance. & Unawareness                         & Privacy policy has not been presented to user. \newline User ignored privacy policy and did not read it. \\ \hline
			Insurance company has access to detailed blood-sugar values.                            & Information Disclosure              & Detailed blood-sugar values are sent to insurance company as part of the general therapy data.  \newline User decides to stop using the device and sends it back to the insurance company without deleting its content. \\ \hline
			Insurance company has access to detailed location data.                                 & Information Disclosure              & Detailed location data is sent to insurance company as part of the general therapy data. \newline User decides to stop using the device and sends it back to the insurance company without deleting its content. \newline High score allows assumptions on health state.  \\ \hline
			Smart device company has access to detailed blood-sugar values.              & Information Disclosure              & Analytics data includes detailed blood-sugar values.\newline User sends device to company for repair without deleting its content.                               \\ \hline
			Smart device company has access to detailed location data.                   & Information Disclosure              & Analytics data includes detailed location data. \newline User sends device to company for repair without deleting its content.                               \\ \hline
			Other players can track location of player.                                         & Information Disclosure              & High scores include location information.                                                                           \\ \hline
			Other players can estimate health state of player.                                      & Information Disclosure, Linkability & High score allows assumptions on health state.                                                                 \\ \hline
			Other players can see identity (name, address) of player.                               & Identifiability, Unawareness        & High scores include personal information of player.                                                            \\ \hline
			Physician receives detailed location information.                                          & Information Disclosure              & Long-term health information includes location data.                                                           \\ \hline
			Parents can track location of children.                                                 & Information Disclosure              & Parent alert system always provides location information.         \\ \hline
			Urban planners can identify player from provided gps- and health-data.                         & Identifiability & Submitted data sets include information about player. \newline Submitted data sets include pattern, that can identify player.      \\ \hline
			Urban planners can link individual data sets so they know that they come from the same player. & Linkability     & Submitted data sets include information about player. \newline Submitted data sets include pattern, that can identify individuals. \\
		\end{tabularx}
		\caption{Adverse consequences and their vulnerabilities in our eHealth scenario. Their LINDDUN category shows by which kind of privacy threat they are caused.}
		\label{tab:AdverseConsequences}
	\end{table*}
	
	\subsubsection{Link Adverse Consequences to Vulnerabilities}
	
	As already inferable from the \cref{tab:AdverseConsequences} the vulnerable system states should be linked to all the adverse consequences that can be caused by them. This has to be done iteratively for each adverse consequence.
	
	\subsubsection{Specify Privacy Constraints}
	
	\textbf{Privacy constraints} ensure that vulnerable system states do not occur. They are created basically by negation of the vulnerabilities. As an example, the vulnerability \emph{General therapy data includes detailed blood sugar values} can be converted to the privacy constraint \emph{Exported therapy information must not include detailed blood sugar values}. (Here we should be more specific what \emph{detailed} means, but for this exploration it should be enough).
	
	If we make sure all the privacy constraints are enforced/followed correctly then the previously defined adverse consequences are prevented. This is why we want to find control actions that could violate these constraints in the following steps.
	
	\subsubsection{Specify Design Requirements}
	This step is skipped for this exploration.
	
	\subsubsection{Create Control Structure Model}
	An important part of STPA is the system \textbf{Control Structure Model}. It contains the \emph{processes} that are to be controlled using a so called \emph{feedback-loop}. The feedback loop is created by attaching a \emph{sensor} that checks the process and reports to a \emph{controller}. The controller then evaluates the sensor value and uses an \emph{actuator} to control the process again. The terminology originates from the safety anaylsis where the system is build up of these parts. It is not yet defined how these elements have to be used in the privacy analysis.
	When analyzing existing systems we can eventually use their existing control structure diagram in this step. If a new system is analyzed we have to create a control structure diagram.
	
\begin{figure*}[t]
		\includegraphics[width=\textwidth]{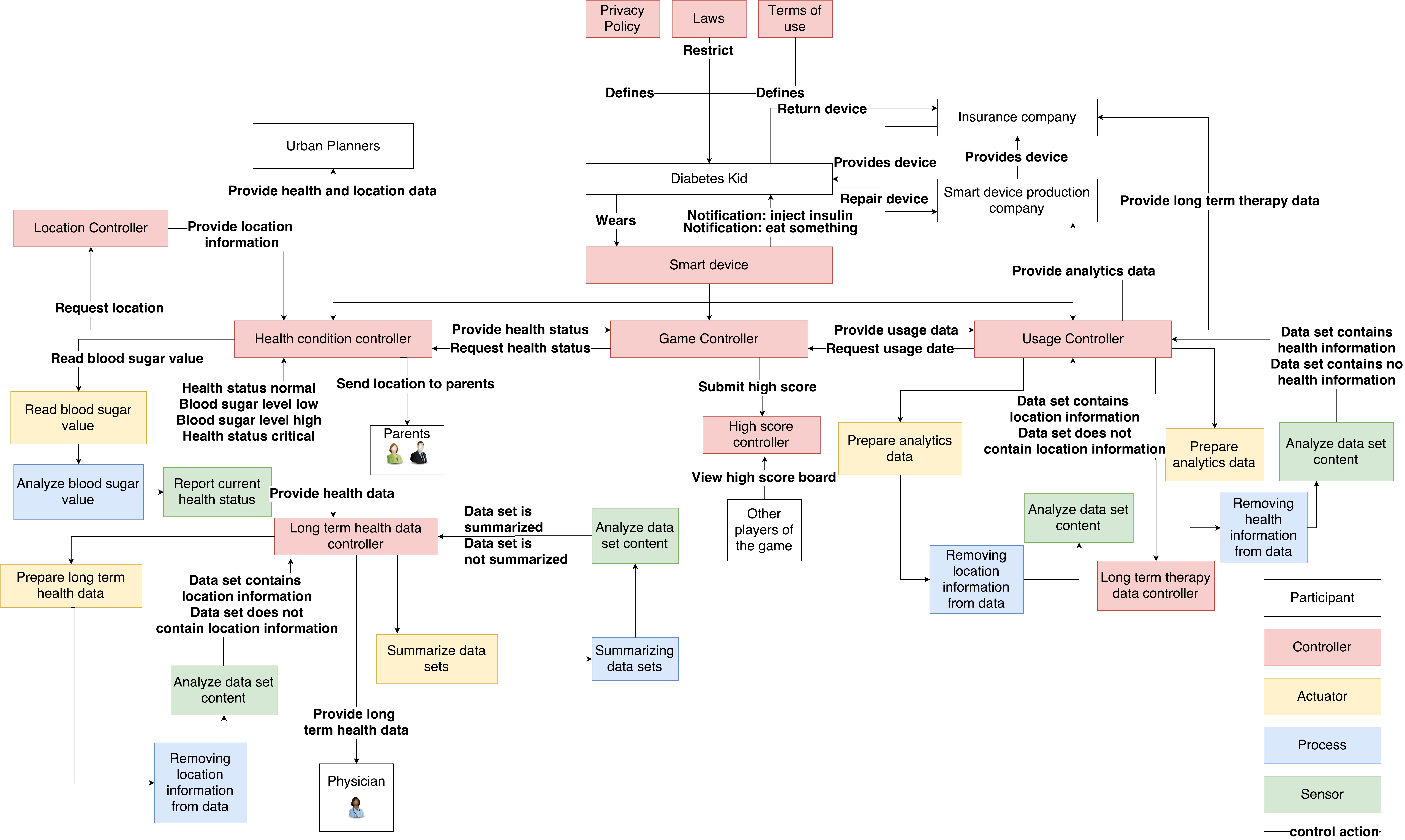}
		\caption{The system with its control structure in a diagram after several iterations. The patient itself, their parents, the smart device producing company, the physician, the insurance company and other players of the game. Different controllers within the watch ensure that data sets are only redirected to specific participants if appropriate requirements are met.}
		\label{fig:eHealthControlStructure}
	\end{figure*}

	The control structure diagram of our scenario is depicted in \cref{fig:eHealthControlStructure}. There are many involved parties with different relationships and interests: The \emph{child} plays an important role; it uses the \emph{smart device}. The smart device is capable of \emph{measuring the blood sugar level} and can \emph{locate its position} using the GPS. Whenever a blood sugar change occurs, the \emph{blood sugar controller} is triggered. This controller decides whether an action is necessary: It could notify the \emph{game controller} to motivate the patient to inject insulin in exchange for in-game rewards, and it can notify the \emph{parent alert controller} in case of an extreme blood sugar value to get help for the child. The game controller also includes a \emph{high score controller} which can \emph{share scores} with other players of the game. \emph{Physicians} can access long-time measurements to be able to discuss and improve the therapy. Analytics data, usage data and crash reports of the smart device are sent out to the \emph{smart device manufacturer} to improve their service or the device. The \emph{health insurance company} is interested in general usage data to be able to see if participants are using the smart device on a regularly basis and correctly. Using this technology more often leads to better insulin injection results and a more stable health condition of the child. This decreases the expenses of the insurance for a specific patient and can therefore decrease their insurance contribution. The creation of this control structure is not easy and requires a lot of domain knowledge. Also it requires decisions on where the system boundaries are and how detailed components are modeled.
	
	\subsection*{Step 1: Control Actions}
	\setcounter{subsubsection}{0}
	\subsubsection{Derive Control Actions}
	The created control structure diagram is used as a basis to derive the existing control actions. They can be taken directly from the model.
	
	\subsubsection{Define Privacy-Compromising Control Actions and  3) Specify Corresponding Privacy Constraints}\stepcounter{subsubsection}
	More important are the \textbf{privacy-compromising control actions} which violate privacy constraints when being executed. The goal of this step is to find all privacy-compromising control actions. In the end, these privacy-compromising control actions are the flaw of our system and need to be tamed by the engineers.  Each privacy constraint is enforced by a controlling component (see \cref{fig:eHealthControlStructure}). According to STPA-Priv, privacy-compromising control actions can be classified in one of the following four categories:
	\begin{enumerate*}[label=\arabic*.]
		\item Not providing the control action, when it should be provided.
		\item Providing the control action, when it should not be provided.
		\item Providing the control action too early, too late or in wrong order.
		\item Stopping the control action or applying it too long.
	\end{enumerate*}
	Yet we feel unsure whether these four categories apply for privacy. Still, it depends on how the control actions are modeled and/or actually executed.
	
	When looking at the privacy constraints, we have to find the appropriate control actions from the control structure in \cref{fig:eHealthControlStructure} that is responsible for ensuring the privacy constraint.
	As an example, the control action \textit{send analytics data} is responsible for ensuring that \emph{the privacy policy has been presented to the user}, that \emph{the user must read the privacy policy}, that \emph{analytics data must not include location information} and that \emph{analytics data must not include blood sugar values}. These cases are then listed in their appropriate category, caused by the privacy-compromising control action \emph{send analytics data}. This results in vulnerabilities like \emph{sending analytics data causes vulnerability when user is not aware of analytics program} or \emph{sending analytics data causes vulnerability when data includes blood sugar information}. 
	
	\subsection*{Step 2: Causal Analysis - Derivation of Causal Factors}
	The previous step generated a list of privacy-compromising control actions that can violate privacy constraints and therefore potentially cause vulnerable system states. They describe \emph{what} could go wrong. The last step of STPA-Priv concludes scenarios that describe \emph{how} a privacy-compromising control action might be executed. This is not limited to simple components, but can occur in conjunction with components and control actions within the whole socio-technical system. This is also often referred to as \emph{worst case scenario}. We have to look at vulnerabilities that are caused by these control actions to find causal scenarios. For example he control action \emph{Send analytics data} can cause different vulnerable states, such as \emph{sending analytics data when user is not aware of analytics program}. This can happen when \emph{the user did not read the privacy policy, or the agreement has not been made available to the user}. These are then the first causal scenarios. The next vulnerability is \emph{providing analytics data when data includes detailed blood sugar values}. This can be caused by a scenario in which \emph{the usage controller filters data incorrectly}. 
	
	Control actions that can be referred to a causal scenario require a risk management response. This can be a privacy policy or terms of use that all parties and components must comply to or other kinds of actions. These are not part of the analysis method and are of course system specific.
	
	\section{Discussion and Conclusion}
	This was a very brief description of the application of STPA-Priv to a complex eHealth scenario. 
	
	The application of STPA-Priv is straightforward. There are a lot of steps which build up on each other, but they follow a simple logic: Define privacy parameters (like adverse consequences), find vulnerabilities, negate vulnerabilities to create constraints, and then find control actions that violate these constraints. During the application we often recognized that we might have to add some things to previous steps and then step through all following steps again. This was no problem so STPA-Priv can and should also be applied iteratively if knowledge about the system is gained during or after the application.
	Despite this simple logic we found the application not easy. This is because to understand what we have to do in each step, a lot of additional information about the process has to be gathered. This information, for example how to choose a threat catalog or how to create the control structure model, is currently not included in the available literature. The most convenient place would be if the XSTAMPP tool would offer this information to the user e. g. in form of a tutorial for each step. Anyway a formal documentation of the method would be a good start. Currently XSTAMPP offers for the steps that require listing artifacts (like adverse consequences or control actions) only very humble tabular input forms. Nevertheless, a nice feature of XSTAMPP is the visual editor for the control structure model. We know that for the safety analysis this control structure representation can already been used to formally validate the model against the safety constraints. But this is currently not tailored for the privacy analysis. Also, it misses vital information on how the control structure model components have to be used in a privacy analysis. 
	
	\textbf{Next steps:} We want to compare STPA-Priv to the LINDDUN privacy threat modeling to find out if one method finds privacy risk that the other method does not and/or if one method requires less effort and resources. Also, we want to create formal documentation for STPA-Priv and make XSTAMPP more self-contained by adding its own privacy threat catalog.

	\section*{Acknowledgment}
	This work was financed by the Baden-Württemberg Stiftung.

	% References are allowed on only one additional page! ESPRE 2017
	\clearpage
	
	
	\bibliographystyle{IEEEtran}
	\bibliography{database}

\end{document}